\documentclass[12pt,epsf,psfig]{article}
\input{epsf}
\usepackage[dvips]{epsfig}  

\newcommand{\be}{\begin{equation}}
\newcommand{\ee}{\end{equation}}    
\newcommand{\mycaption}{\begin{center}%
Figure\thefigure%
\end{center}
\addtocounter{figure}{1}}

{\end{figure}%
\addtocounter{figure}{1}}

\begin{document}

\begin{center}

{\Large\bf Thin-film effects on the surface stopping power
of a free electron gas}\\ 

\vspace{0.5cm}

{A. Garc\'\i a-Lekue$^{\rm (1,}\footnotemark[1]^{)}$
and
J.M. Pitarke$^{\rm (1,2)}$}\\

\vspace{0.6cm}

{\it $^{\rm (1)}$ Materia Kondentsatuaren Fisika Saila, Zientzi Fakultatea,
Euskal Herriko Unibertsitatea, 644 Posta kutxatila,\\
48080 Bilbo, Basque Country, Spain}\\
{\it $^{\rm (2)}$ Donostia International Physics Center (DIPC) and Centro Mixto
CSIC-UPV/EHU, Donostia, Basque Country, Spain 
}\\

\end{center}

\footnotetext[1]{ Electronic address: wmbgalea@lg.ehu.es; Fax:
+34-94-464-8500}

\begin{abstract}

The electronic properties of thin films present quantum-size effects, 
which are a consequence of the finite size of the system. Here we
focus on the investigation of these effects on the electronic energy loss
of charged particles moving parallel with thin metallic films. The energy
loss is calculated, within linear-response theory, from the knowledge of the 
density-response function of the inhomogeneous electron system, which
we evaluate either in the random-phase approximation or with the use of an
adiabatic and local exchange-correlation kernel. 

\end{abstract}



\section{Introduction}

The interaction of moving ions with solids has represented an active field of
basic and applied physics\cite{exp1,exp2}. Charged particles moving near
metallic surfaces loose energy as a consequence of the creation of different
kind of excitations in the metal, such as electron-hole pairs and both bulk and 
surface plasmons\cite{Pines,Ritchie1}. Also, the theoretical understanding of
the electronic excitations is well known to be relevant in surface physics, as
these modes are invariably involved in a variety of surface
spectroscopies[5-9].

Quantum-size effects (QSE) were first investigated by
Schulte\cite{Schulte} within a jellium model of the thin film and later by
Feibelman\cite{Feibelman}, showing that the various physical properties exhibit
an oscillatory behaviour as a function of the thickness of thin metallic films.
These effects, which decrease as the size of the thin film increases, can be
observed experimentally\cite{expqse}. QSE on the surface energy and the work
function have been examined recently, within a stabilized jellium model of the
electron system\cite{Sarria}.

In this paper we focus on the investigation of QSE on the energy-loss spectra of
charged particles moving parallel with metallic slabs. The energy loss is
calculated, within linear response theory, from the knowledge of the 
density-response function of the inhomogeneous electron system, which we
evaluate either in the random-phase approximation or with the use of an
adiabatic and local exchange-correlation kernel. In Section 2 we
present general expressions for the energy loss of charged particles moving
along a definite trajectory. The results of our self-consistent calculations
are presented in Section 3, and in Section 4 our main conclusions are
summarized. Atomic units are used throughout, i.e., $e^2=\hbar=m_e=1$.

\section{Theory}

We consider a recoiless particle of charge $Z_1$ moving with constant velocity
${\bf v}$ along a definite trajectory at a fixed distance $z$ from the
planar surface of a bounded three-dimensional electron gas that is
translationally invariant in two directions. The energy that the moving
particle looses per unit path length, i.e., the so-called stopping power of the
electron system may be obtained from the gradient along the particle trajectory
of the self-consistent potential set up by the particle in the electron
system and evaluated at the position of the particle\cite{Flores}. Within linear
response theory, one finds\cite{aran}: 
\begin{equation}\label{stop}
-{dE\over dx}=-{2\over v}\,Z_1^2\int{d{\bf
q}_\parallel\over(2\pi)^2}\int_0^\infty d\omega\,\omega\,
{\rm Im}W(z,z;{\bf q}_\parallel,\omega)\,\delta(\omega-{\bf
q}_\parallel\cdot{\bf v}),
\end{equation}     
where ${\bf q}_\parallel$ is the momentum transfer in the plane of the
surface, $\omega$ represents the energy transfer, and $W(z,z;{\bf
q}_\parallel,\omega)$ is the screened interaction
\begin{equation}\label{screened2}
W(z,z';q_\parallel,\omega)=v(z,z';q_\parallel)
+\int dz_1\int dz_2\,
v(z,z_1;q_\parallel)\,\chi(z_1,z_2;q_\parallel,\omega)\,v(z_2,z';q_\parallel),
\end{equation}
$v(z,z';q_\parallel)$ and $\chi(z,z';q_\parallel,\omega)$ being two-dimensional
Fourier transforms of the bare Coulomb potential and the density-response
function, respectively\cite{Pines2}.

The stopping power of the electron system can be described by means of
$P(\omega)$, the total probability for the moving particle to exchange energy
$\omega$ with the medium:
\begin{equation}
-{dE\over dx}={1\over v}\int_0^\infty d\omega\,\omega\,P(\omega),
\end{equation}
where
\begin{equation}\label{ptwo}
P(\omega)=-{Z_1^2\over\pi^2 v}\int_0^\infty dq_x\,
{\rm Im}W(z,z;q_\parallel,\omega),
\end{equation}
with $q_\parallel=\sqrt{q_x^2+(\omega/v)^2}$ and $q_x$ being the momentum
transfer along the particle trajectory.       

The key ingredient of our energy-loss calculations is the density-response
function of the electron system, which is known to satisfy the following
integral equation\cite{tddft}
\begin{eqnarray}\label{chi}
\chi(z,z';q_\parallel,\omega)=&&
\chi^0(z,z';q_\parallel,\omega)+\int dz_1\int{\rm d}z_2\, 
\chi^0(z,z';q_\parallel,\omega)\cr
&&\times\left[v(z_1,z_2;q_\parallel)+f_{xc}(z_1,z_2;q_\parallel,\omega)\right]
\,\chi(z_2,z';q_\parallel,\omega),
\end{eqnarray}
where $\chi^0(z,z_1;q_\parallel,\omega)$ is the density-response function
of non-interacting electrons moving in the effective Kohn-Sham potential of
density-functional theory (DFT)\cite{kohn65}, and the
kernel $f_{xc}(z,z';q_\parallel,\omega)$ accounts for exchange-correlation (xc)
effects beyond a time-dependent Hartree approximation.

Exchange-correlation effects are usually introduced within the local-density
approximation (LDA) of DFT, by replacing the xc potential at $z$ by that of a
uniform electron gas of density  $n(z)$. The xc kernel entering Eq. (\ref{chi})
is then set either equal to zero [this is the random-phase approximation (RPA)]
or equal to the static ($\omega=0$) xc kernel
\begin{equation}\label{xckernel}
f_{xc}^{ALDA}(z,z';q_\parallel,\omega)=\left[{dv_{xc}(n)\over
dn}\right]_{n=n(z)}\,\delta(z-z').
\end{equation}
This is the so-called adiabatic local-density approximation (ALDA).

We consider a jellium slab of thickness $a$ normal to the $z$ axis, 
consisting of a fixed uniform positive background of density:
\begin{equation}
n_+(z)=\cases{\bar n,&$-a\leq 0$\cr\cr 0,& elsewhere,}     
\end{equation}
plus a neutralizing cloud of interacting electrons of density $n(z)$. The
positive-background charge density is $\bar n=q_F^3/3\pi^2$, where 
$q_F=(9\pi/4)^{1/3}/r_s$ is the Fermi momentum and $r_s$ is the Wigner-Seitz
radius.

To compute $\chi(z,z';q_\parallel,\omega)$, we follow the method described in
Ref. \cite{Eguiluz}. We first assume that $n(z)$ vanishes at a distance $z_0$
from either jellium edge, and expand the one-electron wave functions in a
Fourier sine series. The distance $z_0$ and the number of sine functions kept in
the expansion of the wave functions are chosen sufficiently large for our
calculations to be insensitive to the precise values employed. We then 
introduce a double-cosine representation for the density-response function,
and find explicit expressions for the screened interaction and the energy-loss
probability in terms of the Fourier coefficients of the density-response
function\cite{aran}.

QSE are originated in the quantization of the energy levels normal to the
surface. As the slab-thickness $a$ increases new subbands for the $z$
motion become occupied, thereby leading to oscillatory functions of $a$
with and amplitude that decays approximately linearly with $a$ and a
period that equals $\lambda_F/2$, $\lambda_F=2\pi /q_F$ being the
Fermi wavelength. These size effects are responsible for the oscillatory
behaviour of the electronic density induced in the electron system by the
external probe and also for the oscillations of the energy-loss function and
the stopping power with the system size. The results presented below correspond
to slabs with the number $n$ of occupied subbands in the range $n=11-14$, for
which $a\sim 4-7\,\lambda_F$. 

\section{Results}

We have investigated thin-film effects on the energy-loss spectra of aluminum
slabs, for which $r_s=2.07$ and $\lambda_F=6.77\,a_0$ ($a_0$ is the Bohr radius,
$a_0=0.529\,{\rm\AA}$). We set $Z_1=\pm 1$ and our results can then be used for
arbitrary values of $Z_{1}$, as the energy-loss probability is, within
linear-response theory, proportional to $Z_1^2$.

The main ingredient of our calculations is the energy-loss
function ${\rm Im}W(z,z;q_\parallel,\omega)$. Fig. 1 shows this quantity, as
obtained from Eq. (\ref{screened2}) with use of either the RPA or the ALDA
density-response function, versus the slab-thickness $a$ and for $z=2\,a_0$,
$q_\parallel=0.4\,q_F$, and $\omega=0.2\,\omega_p$ [$\omega_p=(4\pi\bar
n)^{1/2}$ is the classical plasma frequency of a uniform electron gas of
density $\bar n$]. As $a$ increases, new subbands for the $z$ motion become
occupied. For $r_s=2.07$, the $n=11,12,13,14$ subbands fall below the Fermi
level for $a_n=4.95,5.46,5.96,6.46\,\lambda_F$. When a new subband is pulled
below the Fermi level, the parallel Fermi sea built upon the newly occupied
subband acquires more electrons, thereby increasing the screening and
decreasing the energy-loss function. However, this effect is eventually
overcome by the fact that all the subbands for the $z$ motion get deeper with
increasing film thickness. When $a$ is increased by $\sim\lambda_F/2$, a new
subband begins to be filled and a new oscillation begins. The energy-loss
function of a semi-infinite medium can then be extrapolated with
the use of the following relation\cite{refqsefor}
\begin{equation}\label{qsefor}
{\rm Im}W={{\rm Im}W[a_n^-]+{\rm Im}W[a_n]+{\rm Im}W[a_n^+]\over 3}, 
\end{equation}
where $a_n^-=a_n-\lambda_F/4$ and $a_n^+=a_n+\lambda_F/4$.

The impact of short-range xc effects on the energy-loss function is
investigated by employing the many-body kernel of Eq. (\ref{xckernel}). These
effects provoke a reduction in the screening of the electron-electron
interaction, thereby increasing the energy loss for all projectile
trajectories. This is observed in Fig. 2, where the RPA and
ALDA energy-loss function is shown as a function of the $z$ coordinate [$z\ge
0$ in the vacuum, at the right-hand side of the slab] for two different values
of $a$ corresponding to a local minimum ($a=a_{11}$) and one of the two local
maxima about the minimum ($a=a_{11}^-$), and the same values of $q_\parallel$ and
$\omega$ as in Fig. 1. The energy-loss function of the semi-infinite medium is
also shown, by a solid line. We find that size effects become more significant
as the particle moves away from the surface into the vacuum, which is due to the
fact that as $z$ increases the external field couples mainly with plasmon
modes. Since these are modes characterized by their long wavelength, their
behaviour depends strongly on the overall size of the system. For particles
moving close to the surface the interaction has a short-range character, as
the excitation of electron-hole pairs plays an increasing role. 

Fig. \ref{fig4} shows the RPA and ALDA probability $P(\omega)$, as obtained from
Eq. (\ref{ptwo}) versus the slab thickness $a$, for an external particle
with $v=0.5\,v_0$ [$v_0$ is the Bohr velocity, $v_0=2.19\times10^6\,{\rm
ms^{-1}}$] and $z=2\,a_0$ to exchange energy $\omega=0.2\,\omega_p$ with an
Al slab. This figure exhibits an oscillatory behaviour similar to that
shown in Fig. 1 for the energy-loss function. In Fig. \ref{fig3} we have
represented the RPA and ALDA energy-loss probability $P(\omega)$, as a function
of the energy transfer $\omega$, with the same values of $v$, $z$ and $a$ as
in Fig. 3. As in the case of the energy-loss function, ALDA probabilities are
well over those obtained in the RPA, due to the presence of short-range xc
effects.

The dependence of the RPA and ALDA stopping power on the thickness $a$ of the
slab, as obtained from Eq. (\ref{stop}), is exhibited in Fig. \ref{fig5}. We have
considered a charged particle moving with velocity $v=2\,v_0$ along a
definite trajectory at a fixed distance $z=2\,a_0$ from the right edge of
an Al film. For these values of $z$ and $v$ the ALDA stopping power
is significantly larger than that obtained in the RPA. Fig. \ref{fig32}
shows the RPA and ALDA stopping power, as a function of the velocity, for
a projectile moving at a fixed distance $z=2\,a_0$ from the surface into the
vacuum. As the velocity increases the energy-loss spectrum  of charged particles
moving outside the solid is dominated by long-wavelength excitations and
short-range xc effects, not included in the RPA, tend to become less important.

\section{Conclusions}

We have investigated quantum-size effects on the energy-loss spectra of charged
particles moving parallel with metallic slabs, in the framework of
linear-response theory.

We have found that the quantization of the energy levels normal to the
surface yields a neat oscillatory behaviour of the screened interaction, the
energy-loss probability and the stopping power, as a function of the
thickness of the slab. The amplitude of these oscillations is found to decay
approximately linearly with the slab thickness and their period is found
to equal $\sim\lambda_F/2$. We have also presented self-consistent
calculations for a semi-infinite medium, which have been obtained
from finte-slab calculations with the use of the extrapolation formula given in
Ref. 20.

Our results indicate that size effects become more significant as the particle
moves away from the surface into the vacuum, where long-wavelength excitations
play an important role. As for the impact of short-range xc effects on the
various magnitudes that characterize the interaction of charged particles with
metallic films, we have found that they increase the energy-loss probability,
due to the reduction that these effects provoke in the screening of
electron-electron interactions.

A more detailed presentation of our self-consistent calculations of the
energy-loss spectra of charged particles moving near a semi-infinite electron
system will be published elsewhere\cite{aran}.

\section{Acknowledgments} 

We acknowledge partial support by the University of the Basque Country, the
Basque Unibertsitate eta Ikerketa Saila, and the Spanish Ministerio de
Educaci\'on y Cultura.


\begin{figure}[hbt!]\label{fig1}  
\caption{The RPA and ALDA energy-loss function ${\rm
Im}W(z,z;q_\parallel,\omega)$, as a function of the slab thickness $a$, for
$q_\parallel=0.4\,q_F$, $\omega=0.2\,\omega_p$, and $z=2\,a_0$. Dashed lines
represent the infinite-width limit, as obtained from Eq. (\ref{qsefor}). The
damping parameter in the evaluation of the density-response function of
non-interacting Kohn-Sham electrons has been taken to be $\eta=\omega_p/10$.}
\end{figure}        

\begin{figure}[hbt!]\label{fig2}
\caption{The RPA and ALDA energy-loss function ${\rm
Im}W(z,z;q_\parallel,\omega)$, as a function of the $z$ coordinate, for two
different values of the slab width and the same values of $q_\parallel$,
$\omega$ and $\eta$ as in Figure 1. Thick solid lines represent
the infinite-width limit of Eq. (\ref{qsefor}). Dashed and thin-solid
lines represent the energy-loss function for $a=a_{11}$ and $a=a_{11}^-$,
respectively.}  
\end{figure}   

\begin{figure}[hbt!]\label{fig4} 
\caption{The RPA and ALDA energy-loss probability $P(\omega)$, as a function of
the slab thickness $a$, for $\omega=0.2\,\omega_p$, $z=2\,a_0$, $v=0.5\,v_0$,
and $\eta=\omega_p/10$. Dashed lines represent the infinite-width
limit, which is obtained as in Eq. (\ref{qsefor}) with $P$ instead of ${\rm
Im}W$.}     
\end{figure}    

\begin{figure}[hbt!]\label{fig3}
\caption{The RPA and ALDA energy-loss probability $P(\omega)$, as a function of
the energy transfer $\omega$, for two different values of the slab width and the
same values of $z$, $v$ and $\eta$ as in Figure \ref{fig4}. Thick solid lines
represent the infinite-width limit. Dashed and thin-solid
lines represent the energy-loss function for $a=a_{11}$ and $a=a_{11}^-$,
respectively.} 
\end{figure}   

\begin{figure}[hbt!]\label{fig5} 
\caption{The RPA and ALDA stopping power, $-(dE/dx)$, as a function of the slab 
thickness $a$, for $z=2\,a_0$, $v=0.5\,v_0$, and $\eta=\omega_p/10$. Dashed lines
represent the infinite-width limit, which is obtained as in Eq. (\ref{qsefor})
with $-(dE/dx)$ instead of ${\rm Im}W$.}  
\end{figure} 

\begin{figure}[hbt!]\label{fig32} 
\caption{The RPA and ALDA stopping power, $-(dE/dx)$, as a function of the
velocity $v$, for two different values of the slab width and the
same values of $z$ and $\eta$ as in Figure \ref{fig5}. Thick solid lines
represent the infinite-width limit. Dashed and thin-solid
lines represent the energy-loss function for $a=a_{11}$ and $a=a_{11}^-$,
respectively.}  
\end{figure}

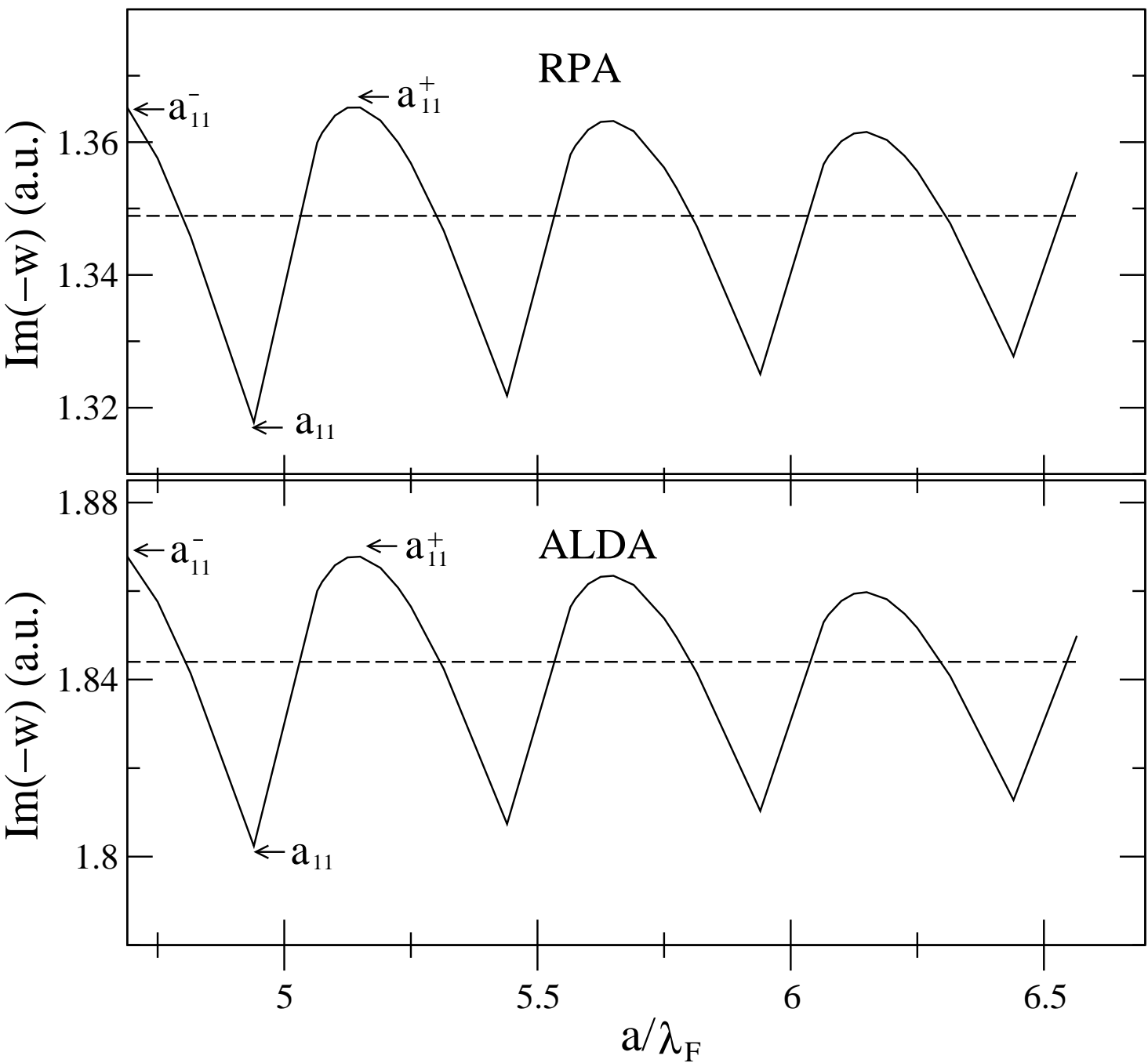
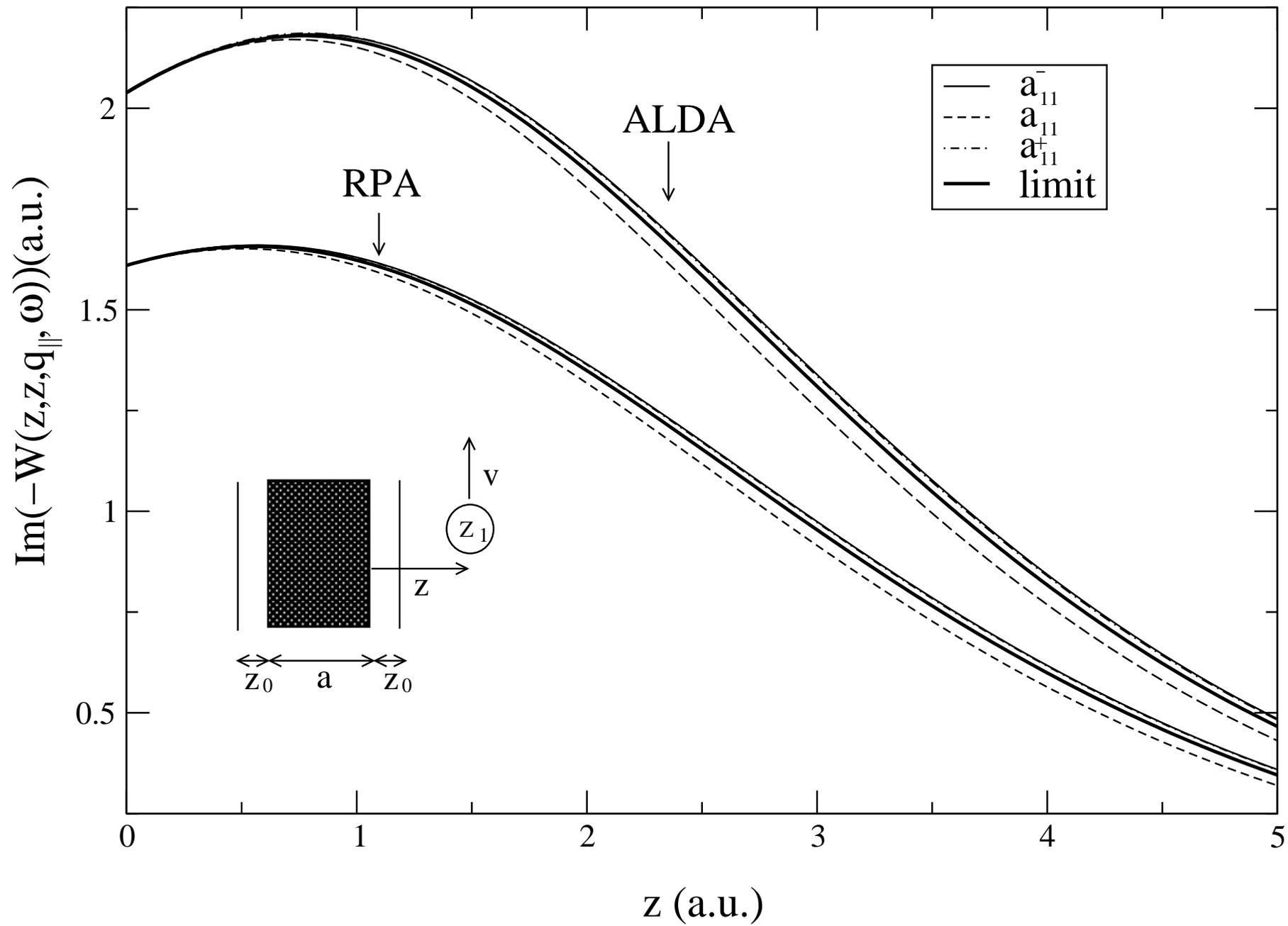
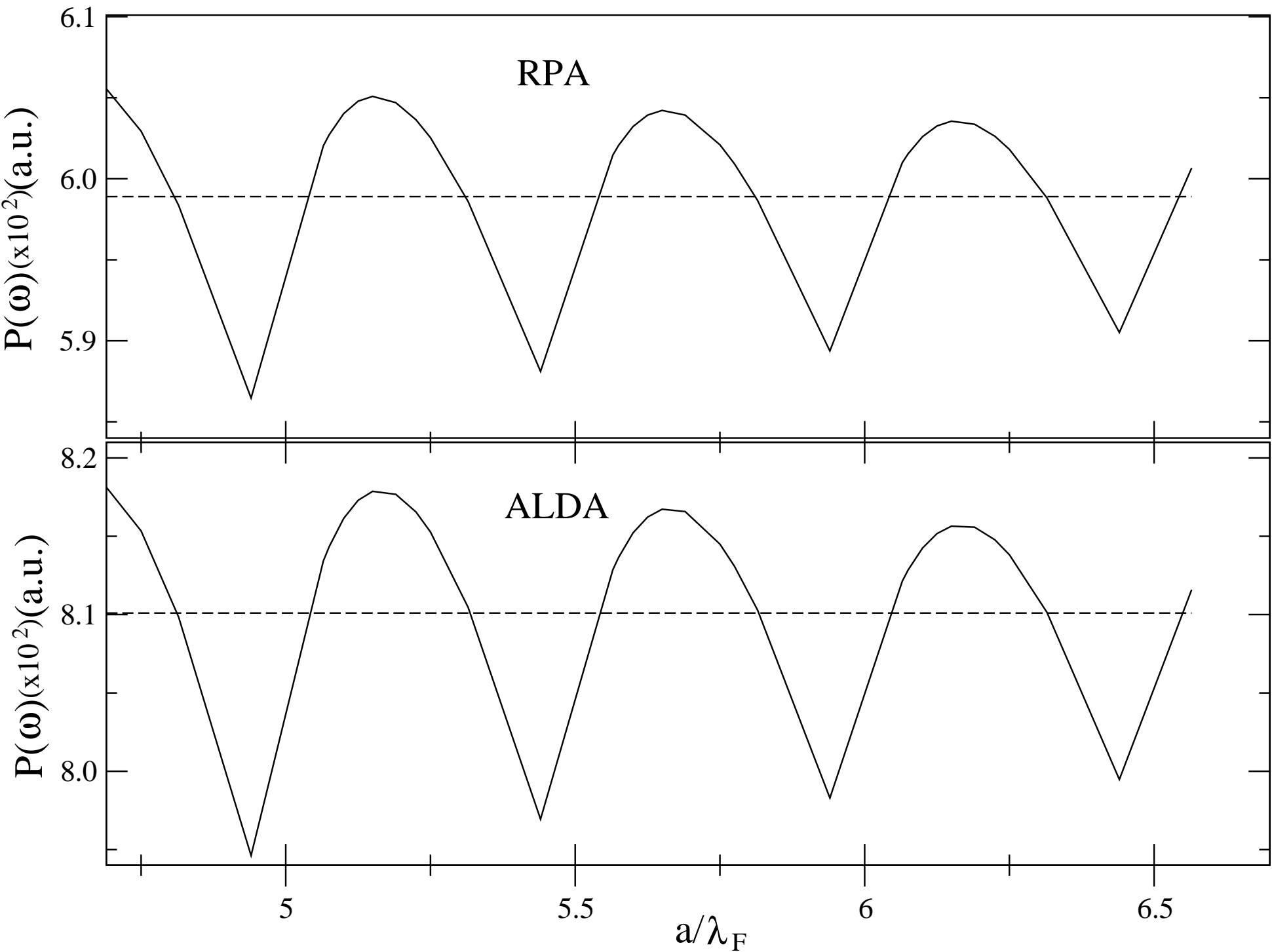
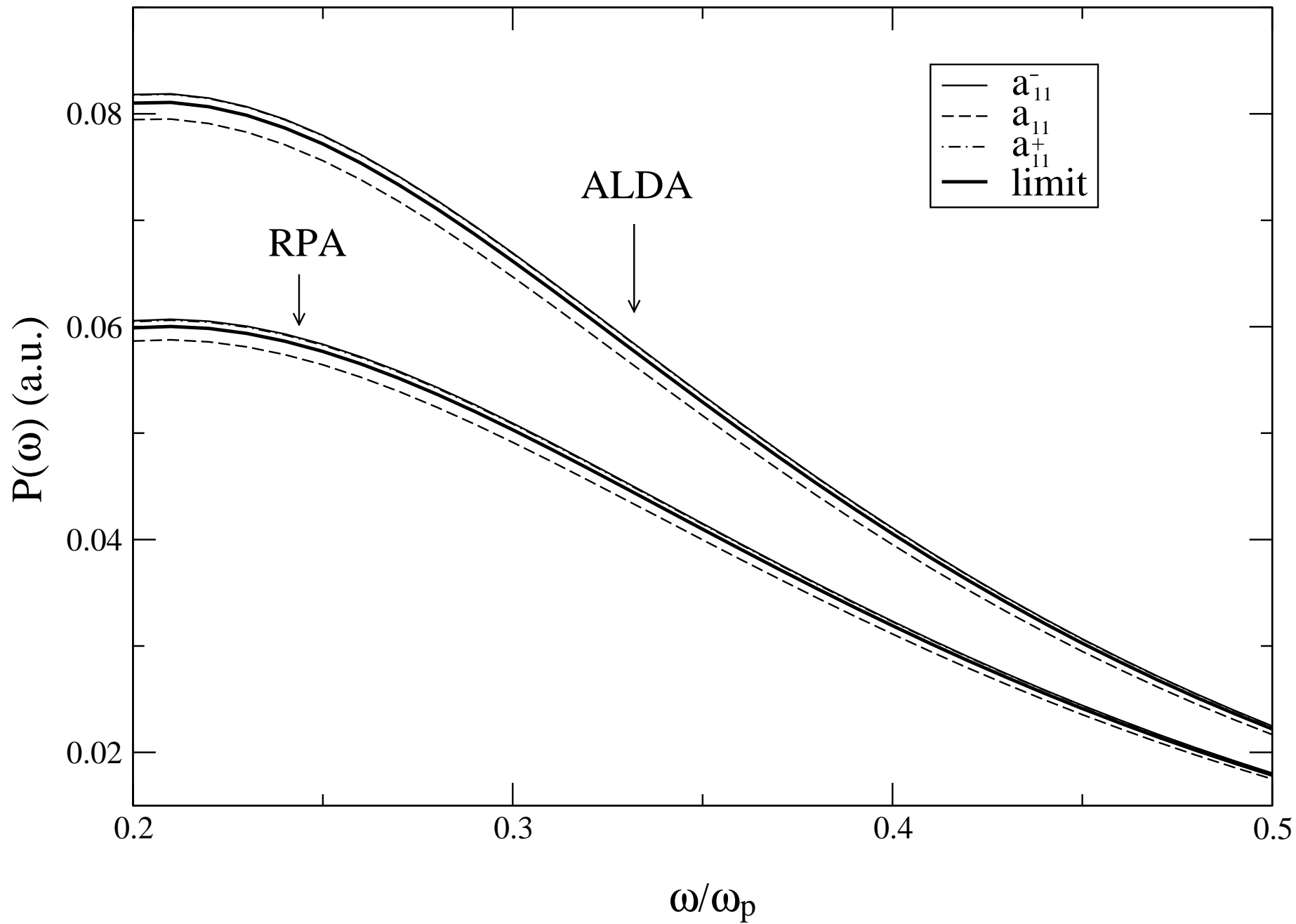
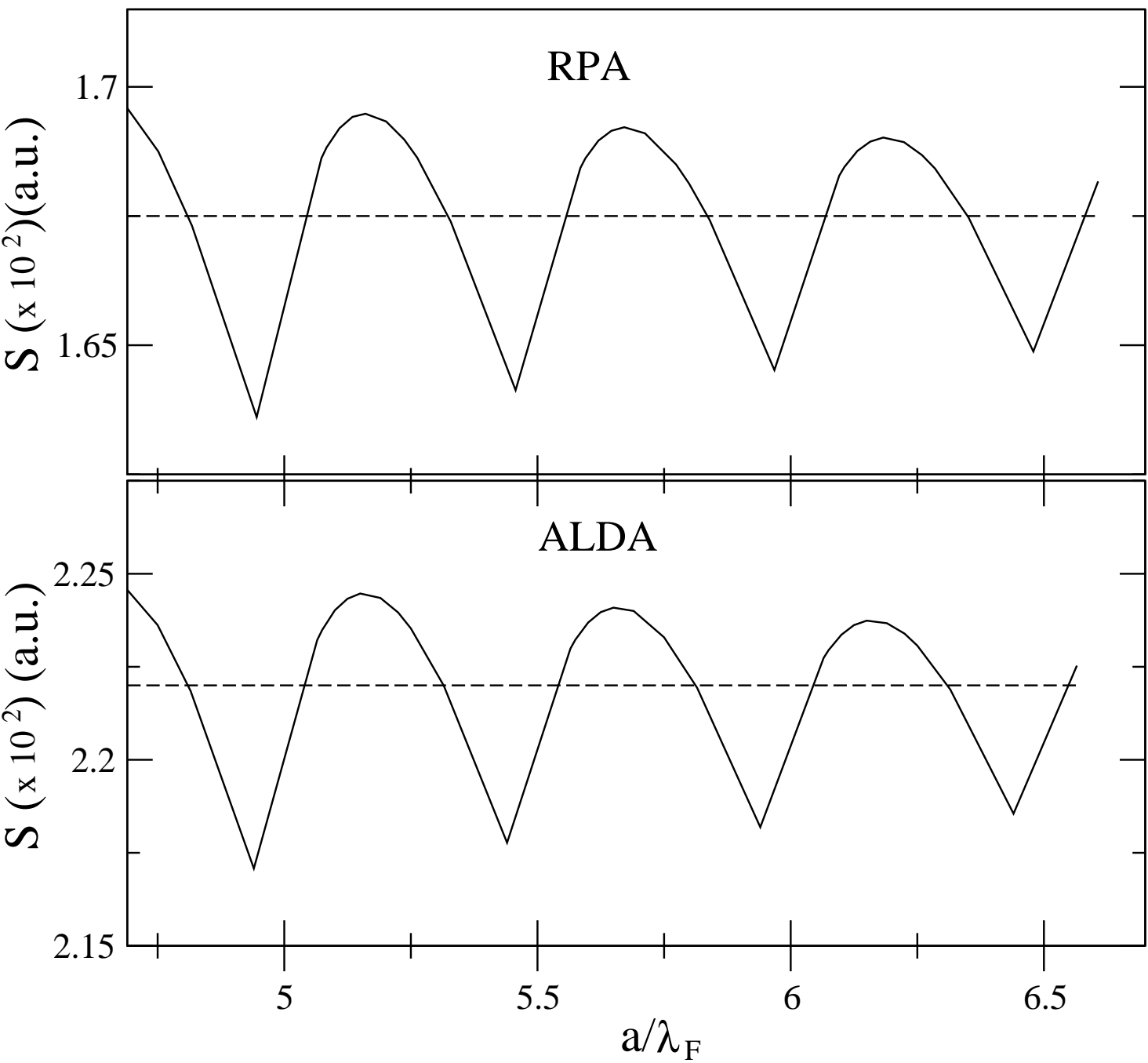
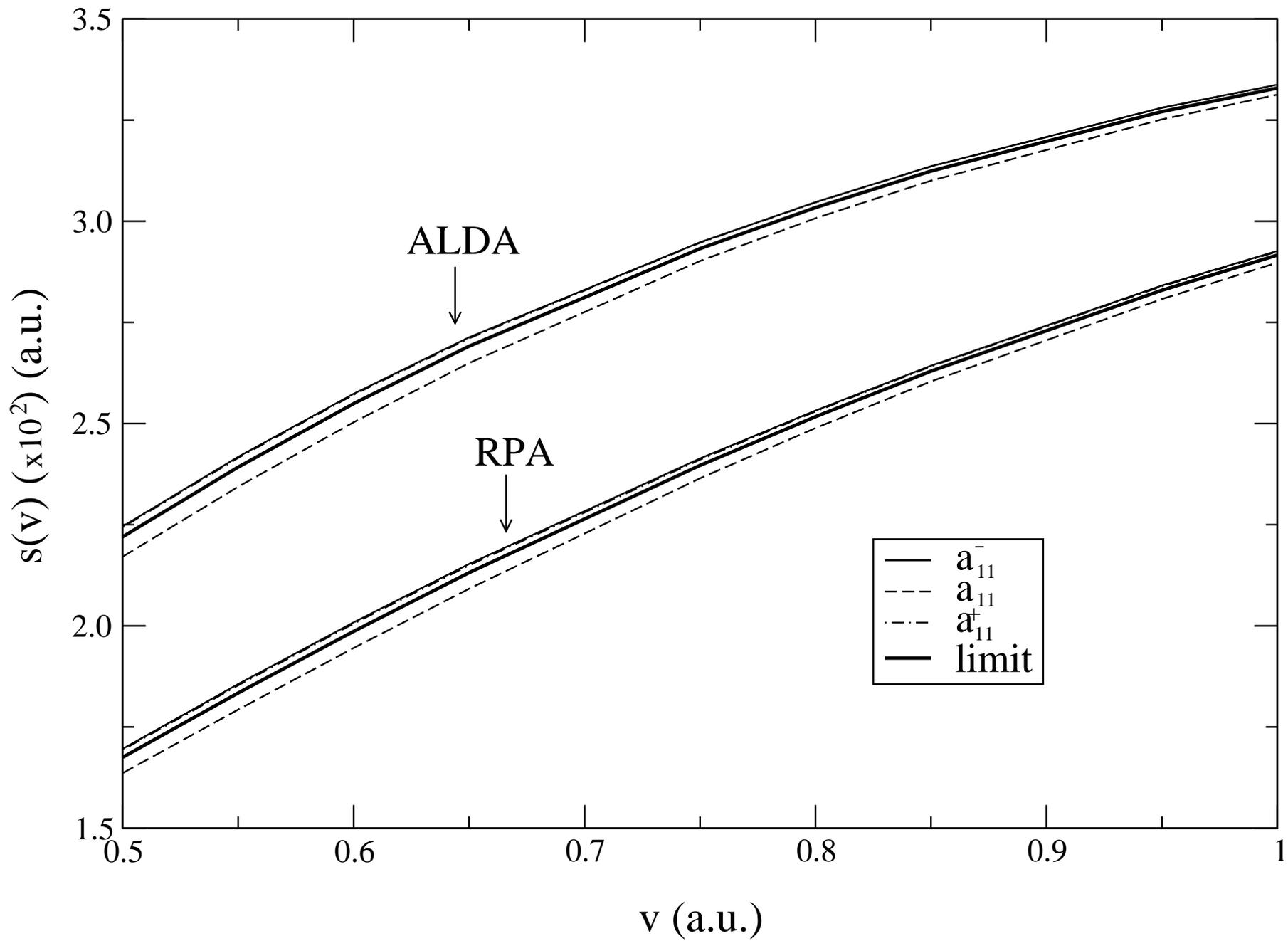
\end{document}